# Novae -The study of the reactive flow

S A Glasner [1] and J W Truran [2,3]
[1]Racah Institute of Physics, The Hebrew University, Jerusalem, ISRAEL
[2]Department of Astronomy and Astrophysics, Enrico Fermi Institute, University of Chicago, Chicago, IL 60637, USA
[3]Physics Division, Argonne National Laboratory, Argonne, IL 60439, USA

E-mail: glasner@phys.huji.ac.il

**Abstract.** There is a wide consensus in the astrophysics community that the mechanism underlying the observed Classical Nova eruptions is a surface thermonuclear runaway. We start this short review with the main observational facts that lead to the theoretical model of a thermonuclear runaway that takes place in an accreted hydrogen rich envelope placed on top of a cool degenerate core of a white dwarf. According to the theory, the accreted envelope becomes unstable to convection days to weeks prior to the runaway. During the extreme stages of the runaway itself, when the burning is most efficient, the envelope is fully convective. Therefore, the elements processed under such extreme conditions are lifted to the outermost regions of the star. A significant fraction of the envelope is ejected during the outburst. The complicated combination of hydrodynamic instabilities and explosive hydrogen burning, close to the surface of the star, gives us a unique opportunity to study this complex reactive flow. The range of core masses, core temperatures and accretion rates introduce a whole range of burning temperatures and densities. Following the description of the "standard" cases, we then focus on rare, but still possible, portions of the relevant parameter space, in which "breakout" of the traditional CNO cycle can occur and lead to heavy element enrichment patterns caused only by breakout burning. We conclude our review with the main challenges that nova theorists face today, with special emphasis on problems related to the nucleosynthesis issues.

## 1. Introduction-Classical novae the observational constraints

Classical Novae are extremely bright outbursts. A rise to peak brightness is followed by steady decline. The time taken for a nova to decay by 2 magnitudes from maximum optical brightness is used to classify a nova by its speed class. A fast nova will typically take less than 25 days to decay by 2 magnitudes and a slow nova can take over 80 days. The bolometric luminosity stays almost constant, at the level of a few $10^4$ solar luminosities ($L_\odot$) for a period that extends from a few months to a few years [1]. In spite of their violence, the amount of material ejected in a nova outburst is usually only about $10^{-6}$-$10^{-4}$ of a solar mass ($M_\odot$). Matter is ejected at velocities as high as several thousand kilometers per second—higher for fast novae than for slow novae. Spectroscopic observations of nova ejecta show that they can be enriched relative to solar abundances in helium, carbon, nitrogen, oxygen, neon, magnesium and even heavier elements. Classical novae eruptions occur on Cataclysmic Variable stars (CV) that consist of binary star systems in which the primary, a white dwarf, accretes solar abundant matter from a secondary, which is a low mass main sequence star. The entire binary system is usually the size of the Earth-Moon system – with an orbital period of 1 to 10 hours. A much broader review with further details can be found in [2] and references therein.

## 2. The "Standard Model" for Classical Novae

The standard model for classical novae consists of a thermonuclear runaway (TNR) occurring in a degenerate hydrogen rich (solar abundance) envelope accreted on a carbon-oxygen (CO) or oxygen-neon-magnesium (ONeMg) white dwarf in a close binary system. As long as the accretion process continues, the degenerate matter that accumulates at the base of the hydrogen rich envelope is compressed and heated. Under the prevailing degenerate conditions, the heated matter is burning without hydrodynamic regulation by expansion. Therefore, an explosive runaway occurs only once the relevant timescale for heat release by hydrogen burning in the degenerate envelope becomes shorter than the typical cooling time by any of the available cooling mechanisms. The runaway phase ends when the energy produced by the burning lifts degeneracy and the pressure decreases. The runaway time is much shorter than the burning time, and therefore only a few percent of the fuel is consumed at this stage. At later stages, the burning sets into a quasi equilibrium. In the quasi equilibrium phase, the burning rate is approximately equal to the outward luminosity. Many 1D hydrodynamic models for hydrogen TNRs on white dwarfs give very good agreement between theory and the main features of the observed phenomenon ([3]-[14]).

## 3. The reactive flow -mixing

As stated above, observations show that for nova outbursts the accreted mass is of solar abundance whereas the ejected mass is enriched by CNO and heavier elements. Since the timescales and the thermodynamic conditions (density, temperature) during the runaway do not predict any significant production of CNO elements, we conclude that there must be a mechanism that mixes the accreted matter with the cold white dwarf core prior to, or during, the runaway. A few mechanisms for such mixing were proposed up to now. All of them claim to predict the right observational amounts of mixing:

- The diffusion layer mechanism. The base of this layer is at the deepest point where hydrogen sinks into the core [15]-[19].
- The shear instability induced by differential rotation during the accretion phase [20]-[25].
- Shear gravity wave breaking on the white dwarf surface [26],[27].
- Undershoot of the convective flow during the runaway [28].

In all the published models, once the temperature at the base of the layer is about 20 million degrees, the CNO burning can't be controlled any more by radiative losses, and the whole envelope becomes convective. In the first three cases mentioned above, at this stage, much before the runaway takes place, the abundance of the whole envelope is totally mixed. In the fourth model, most of the mixing occurs during the runaway itself.

In a series of papers we [29]-[31] studied the convective undershoot mechanism (figure 1). The most significant result of this study is the observed universal behavior of all the models as it is presented in [31]. Multi-D effects give rise to mixing that shortens the rise-time to TNR, and the numerical tests support the conclusion that the overall mixing is at the level of 35-50 % . Mixing in the early stages is small and sums up to about 10%. The fact that at the late stages of the runaway all the models converge to an almost universal model, can be explained by the great sensitivity of the burning rate enhancement by CNO mixing to the temperature. Observations of nova ejecta abundances provide evidence for dredge-up that enriches the envelope with heavy elements up to 30%-40% relative to solar ([32] and references therein). Recently [33],[34] made 2D and 3D detailed surveys that confirm those results.

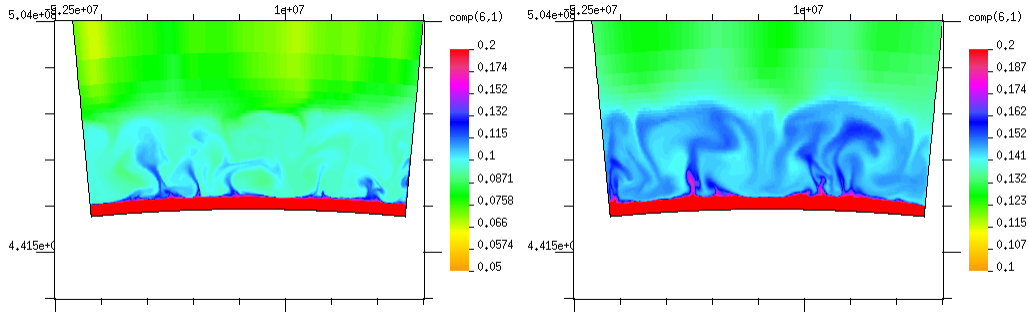

**Figure 1.** Color map of the $^{12}C$ abundance at various stages of the runaway. Left Log(Q)=42 erg/s. Right: Log(Q)= 43 erg/s.

## 4. The reactive flow - unique issues related to the nuclear burning

### 4.1. $\beta^+$ decay time vs. convective turnover time

CNO isotopes unstable to $\beta^+$ decay play a major role in the energetics and nucleosynthesis of the outburst [35]. The four nuclei ($^{13}N$, $^{14}O$, $^{15}O$ and $^{17}F$) have lifetimes of the order of 100-1000 seconds. Once the temperature at the base of the envelope increase above $10^8$ K, the lifetime against proton capture becomes shorter than the $\beta^+$ decay time. Therefore, the abundances of these nuclei increase rapidly. At this stage of the runaway, most of the envelope is convective and the nuclei circulate within the convective cells. As long as no fresh proton capturing CNO elements ($^{12}C$, $^{14}N$, $^{16}O$) appear in the entire convective zone, the energy generation rate is determined by the lifetime of the four $\beta^+$ unstable nuclei and therefore is not temperature dependent. Since the temperature at the upper regions of the convective zone is much lower than $10^8$ K, the only nuclear reactions that occur there are the delayed decays of the $\beta^+$ unstable nuclei. Under these conditions, it is clear that the convective turnover time and the size of the convective cells are the most important parameters impacting the energy generation rate and the nucleosynthesis. After one or two turnover times, the outer part of the envelope expands with very high velocities (about $10^8$ cm/sec). The entropy of the outer parts increases due to the deposition of energy released by the decay of unstable nuclei. Under such conditions the outer zones are detached from the main convective cells underneath.

### 4.2. Limits on energy production rate and undershoot mixing

For any mechanism other than the convective undershoot mechanism, the mixing with fresh $^{12}C$ takes place much before the runaway. Therefore, as was stated above, during the runaway itself, when the convective time-scales are much shorter than the $\beta^+$ decay time virtually all abundant stable CNO isotopes are transformed to $\beta^+$ unstable proton rich isotopes and the burning rate is limited to a constant value related to the total amount of CNO nuclei:

$$q_{max} = 5.8 \times 10^{13} \times (Z_{cno}/0.01) \text{ (erg/gr/sec)}$$

If, on the other hand, there is dredge-up of CNO elements during the runaway itself, there is no such limit on the burning rates. In this case, rates could be much higher and temperature dependent once the rate of ingestion of fresh $^{12}C$ into the convective envelope is high enough. The enhanced burning rate increases the burning temperature and by that can have an effect on the abundances of the processed material. The processed matter is advected outwards by the convective flux and can later be observed in the ejecta. A demonstration of this effect is given in figure 2 and figure 3. The immediate capture of protons and the short period of time for lifting these nuclei to the outermost zones, above the main convective cells, (10-15 seconds) have the following consequences:

- Most of the dredged up $^{12}C$ will end up as $^{12}C$, $^{13}C$ or $^{14}N$.
- Most of the $\beta^+$ products ($^{13}C$ and $^{15}N$) are born in the upper part of the envelope and will not have the chance to capture another proton. Therefore, the $^{13}C/^{12}C$ and the $^{15}N/^{14}N$ ratios are relatively high.
- Novae may contribute significantly to the abundance of the $^{13}C$, $^{15}O$ or $^{17}O$.

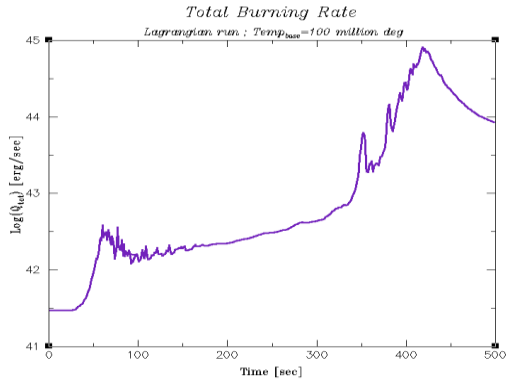
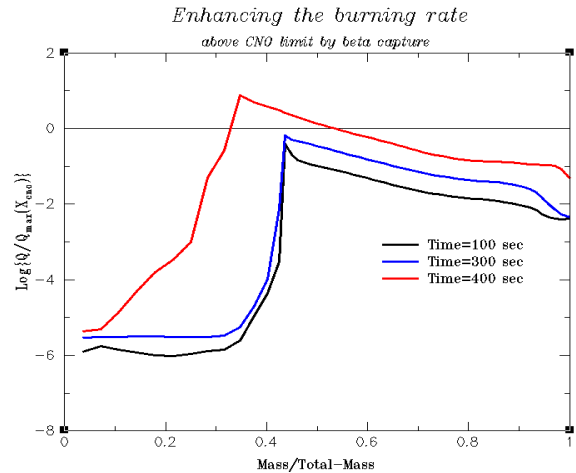

**Figure 2.** The logarithm of the total energy generation rate [ers/sec] vs. time.

**Figure 3.** The ratio of the specific burning rate [erg/gr/sec] to the limiting rate vs. mass at various stages of the runaway (see left figure).

## 5. Burning under extreme conditions – the possibility of breakout from the CNO cycle

### 5.1. Breakout – the relevant parameters

For very slow white dwarf accretors in cataclysmic variables, Townsley&Bildsten [37] found a relation between the accretion rate $\dot{M}$ and the central temperature $T_c$ of the white dwarf. According to this relation, for $\dot{M}$ less than $10^{-10}\ M_\odot/yr$, $T_c$ is much lower than $10^7$K. Motivated by this study, we followed the thermonuclear runaway on massive white dwarfs ($M_{WD}$ = 1.25–1.40 $M_\odot$) with $T_c$ lower than $10^7$K, accreting matter of solar composition. In [38] we demonstrated that in this range of the relevant parameter space, the accreted envelope is extremely massive, and the slope of the relation between the peak temperatures achieved during the runaway and $T_c$ becomes much steeper than its value for $T_c$ above $10^7$K. The peak temperatures we derived were above $5 \cdot 10^8$K and they stay above a critical value for breakout from the conventional "hot carbon–nitrogen–oxygen" cycle ($T_{crit}\sim 4 \cdot 10^8$K) for a few hours. When breakout conditions are achieved, the heavy-element abundances can exhibit a much wider variety of enrichments than what is possible with the common enrichment mechanisms. If indeed observed, such rare novae events will offer a unique opportunity to examine CNO breakout abundances in environments other than X-Ray burst on neutron stars.

In order to ensure that we indeed achieved "breakout" from the conventional "hot CNO" cycle, we conducted a definitive test. We examined a model similar to the nominal model ($M$ = 1.35 $M_\odot$), for which the accreted matter was assumed to include solar concentrations of nuclei only up to fluorine. The energetics of this model were similar to those of the original model. With regard to the nucleosynthesis, the final concentration of iron group elements was determined to be comparable to that of the original model. There is only a small concentration of intermediate-mass elements in the final stages, which confirms that indeed CNO nuclei were burned all the way to iron (figures 4-5).

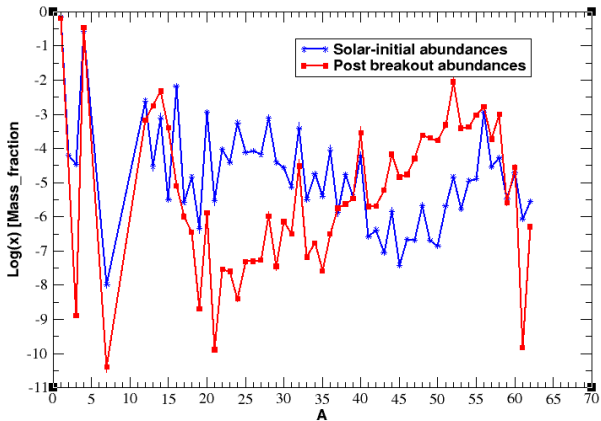
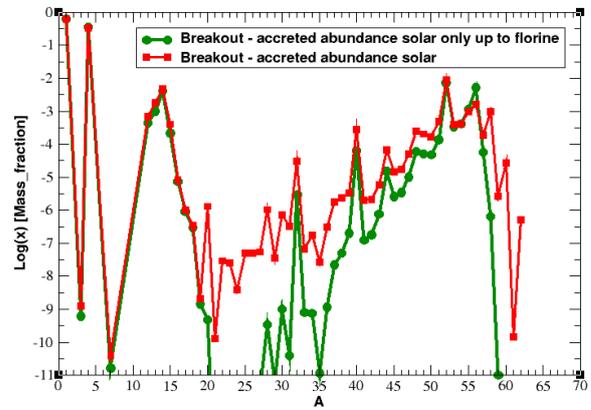

**Figure 4.** Abundances of the elements in the reaction network as a function of the mass number *A*. Presented are both the initial (solar) abundances (blue) and the "final" abundances (red) when the peak temperature has fallen below $T_{crit}$ (M=1.35 $M_\odot$).

**Figure 5.** Abundances of the elements in the reaction network as a function of the mass number *A*. Presented are the "final" abundances for the nominal model (red here and in the left figure) and a model with accreted solar abundances, only up to fluorine (green).

*5.2. Breakout - sensitivity*

For our attempts to demonstrate the sensitivity to the exact value of specific cross sections of the burning reactions, we define two sensitivity levels. The first and major level is a sensitivity that alters the overall energetic of the runaway i.e., the temperature history of the burning, the total burning rate and light curve. We define those reactions as reactions with global sensitivity. In the second case, the sensitivity can show itself only by altering the abundances of specific isotopes without having any substantial effect on the energetics of the runaway. We define those as reactions with local sensitivity. The only global sensitivity we found up to now is for the reaction $\alpha + {}^{15}O \rightarrow {}^{19}Ne + \gamma$. The effect of multiplication of the cross section by a factor of 10 on the overall energy production rate was profound (figure 6).

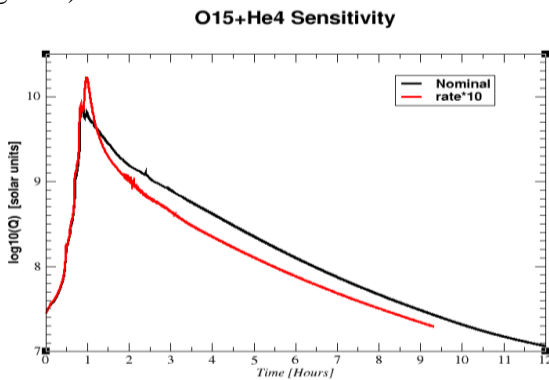

**Figure 6.** The effect of multiplication of the cross section for the reaction $\alpha + {}^{15}O \rightarrow {}^{19}Ne + \gamma$ on the energy production rate.

## 6. Conclusions - Challenges to novae theorists

The main problems facing novae theoretical investigators are the following (the problems related to the nuclear physics are marked in bold letters):

- Modeling of the early luminosity histories of novae in outburst.
- **Identification of the mechanism by which novae envelopes are enriched in heavy elements.**

- Clarifying the consequences of the phase of common envelope evolution that characterizes all novae at maximum light.
- **Systematic interpretation of the observed compositions of novae ejecta.**
- **Understanding of the timescale for post-outburst nova systems to return to minimum.**
- **Gathering accurate data base of abundance observations that is good enough for sensitivity examination of nuclear cross sections.**